\documentclass[aps,pra,numerical,preprint]{revtex4-1}
\usepackage{graphicx}
\usepackage{amsmath}
\usepackage{setspace}
\usepackage{amsfonts}
\usepackage[dvips]{epsfig}
\usepackage{color}

\begin{document}

\title{Application of the spectral element method to the solution of the multichannel Schr\"odinger equation}
\author{Andrea Simoni, Alexandra Viel and Jean-Michel Launay}
\affiliation{Institut de Physique de Rennes, UMR 6251, CNRS and Universit\'e de
Rennes 1, 35042 Rennes Cedex, France}

\date{\today}

\begin{abstract}

We apply the spectral element method to the determination of scattering
and bound states of the multichannel Schr\"odinger equation. In our
approach the reaction coordinate is discretized on a grid of points
whereas the internal coordinates are described by either purely
diabatic or locally diabatic (diabatic-by-sector) bases. Bound levels
and scattering matrix elements are determined with spectral accuracy
using relatively small numbers of points. The scattering problem is
cast as a linear system solved using state-of-the-art sparse matrix
non iterative packages. Boundary conditions can be imposed so
to compute a single column of the matrix solution. A comparison with
log-derivative propagators customarily used in molecular physics is
performed. The same discretization scheme can also be applied to bound
levels that are computed using direct scalable sparse-matrix solvers.

\end{abstract}
%\pacs{34.50.Cx,31.50.Bc}

\maketitle

\section{Introduction} The solution of partial differential equations,
ubiquitous in all areas of physics, can be tackled by a variety of
numerical methods developed over last decades. Solution algorithms
can essentially be divided into propagation and basis expansion
approaches. The former build the solution iteratively from a known
initial value up to the final propagation distance, where suitable
boundary conditions are imposed. Such methods are easy to implement and
cheap in memory storage but provide relatively low convergence rates
as a function of the step size. Large number of steps may therefore
be needed, such that accumulation of roundoff error can limit the
accuracy in particular for complex systems.
Due to its pivotal role in quantum dynamics, the time-independent
Schro\"odinger equations has been granted a particular attention in the
molecular physics community. Popular propagation algorithms include the
log-derivative propagator of Johnson~\cite{1973-BRJ-JCoP-445}, later
improved by Manolopoulos~\cite{manolopoulos1986}, and the renormalized
Numerov algorithm~\cite{1978-BRJ-JCP-4678}.

In basis expansions the solution is determined altogether as a
development over a basis, usually trigonometric or polynomial, with
suitable conditions imposed at the boundaries. A main advantage is
the exponential numerical convergence as a function of the expansion
order~\cite{BK-2001-JPB}. Grid-discretization methods are particular
basis expansions in which each basis functions is nonzero at a unique grid
point. In global approaches the whole interval of interest is represented
as a discrete grid of points. Global grid techniques have been introduced
in molecular physics in the context of the so-called discrete variable
representation (DVR)~\cite{1982-JVL-CPL-483}. One drawback of global
grids is the need to introduce nonlinear coordinate transformations
to efficiently represent complex solutions varying on disparate length
scales such as ultracold processes~\cite{TiesingaPRA98,KokoulineJCP99}.

Local approaches subdivide the interval of interest in subintervals,
often termed elements. Basis functions localized in subintervals are
used to expand the solution. As compared to global approaches, since
the resulting matrix is sparse one can apply performant sparse linear
algebra packages to carry out the operations needed in the specific
problem. Moreover, at least in one dimensional problems the element size
can be tailored to the solution in a straightforward way.

One widespread local approach is the finite element
method~\cite{hughes2000FEM}. Being based on low order polynomial
expansions, such method requires large number of points to achieve
high accuracy. Use of high order polynomials as basis functions in
each element marked the birth of spectral element approaches few
decades ago~\cite{1980-SAO-JCoP-70}. The spectral element method
is nowadays a well established tool in scientific and engineering
computations~\cite{BK-2006-GEK,canuto2007spectral_complex}.

Use of this computational technique in molecular physics has been
pioneered for one dimensional problems in Ref.~\cite{1988-DEM-CPL-23}. More
recently, a spectral element approach in two spatial dimensions has been
presented in~\cite{2000-TNR-PRA-032706,2009-LT-PRA-012719}, where it was
termed finite-element DVR. Appropriate scattering boundary conditions were
imposed using a spatial rotation in the complex plane known as exterior
complex scaling approach~\cite{2000-TNR-PRA-032706,2010-AS-PRA-053845}. A
combination of the spectral element and the slow variable
representation~\cite{1996-OIT-JPB-L389} has been proposed in
order to compute weakly bound states of triatomic systems in the
hyperspherical framework~\cite{2011-HS-JCP-064318}. Finally, the
finite-element DVR has been used as a time-independent representation
in multidimensional time-dependent calculations; See {\it
e.g.}~\cite{2006-BIS-PRE-036708,2007-DAT-PRA-043412,2010-SXH-PRA-056705}.

Main aim of the present work is to explore the usefulness of the spectral
element method in quantum dynamics for time-independent multichannel
problems. More particularly, we point out that combining the spectral
element method with traditional molecular basis or diabatic-by-sector
expansions~\cite{1989-JML-CPL-178} to treat the internal coordinates
optimizes sparsity and size of the discretized Hamiltonian. One major
advantage is that the wavefunction is obtained at no extra computational
cost. Moreover, the spectral nature of the method allows the accuracy of
the solution to be estimated reliably in each region of space. Subsequent
step refinements lead to a grid tailored to the interaction strength in
various regions of space. We show that the spectral element formulation
lends itself to imposing in a natural way different boundary conditions
for scattering and bound-state calculations.

The paper is organized as follows: Sec. II presents the discretization
scheme and introduces various boundary conditions, Sec. III discusses
numerical results on a realistic ro-vibrational system, Sec. IV summarizes
and concludes this work.

% DISCRETIZATION OF THE HAMILTONIAN

\section{Hamiltonian discretization}
\label{sec_discre}

Fundamentals of the spectral-element
approach can be found in textbooks and
articles~\cite{canuto2007spectral_complex,1988-DEM-CPL-23,2000-TNR-PRA-032706}.
In order to set the notation and to illustrate the specific approach
we follow to combine full or locally diabatic expansions and the grid
basis, we reproduce in this section the main steps of the derivation
from scratch.

We consider a generic time-independent multidimensional problem and
identify an \lq\lq external\rq\rq reaction coordinate $R$ describing the
``size" of the system and a set of \lq\lq internal\rq\rq variables denoted
collectively as $\Omega$. For instance, in the two-body problem $R$
typically represents the distance between the particles and $\Omega$ the
orientation of the inter-particle position vector. In three-body systems
$R$ may represent the hyperradius and $\Omega$ a set of hyperangles.
Note that in general $\Omega$ may comprise coordinates with physical
dimension of length, like in the case of our test atom-molecule
ro-vibrational problem described in detail in Sec.~\ref{numerical_tests}.

The time-independent Schr\"odinger equation is schematically written
\begin{equation}
\label{schroedinger}
\left[ -\frac{\hbar^2}{2 \mu}\frac{\partial^2}{\partial R^2}+V(R,\Omega) \right] \Psi(R,\Omega)=E \Psi(R,\Omega) ,
\end{equation}
to be solved in the hyper-region $R_{\text{min}} \leq R \leq R_{\text{max}}$. Here $V$ 
contains various potential energy terms and/or differential operators
acting on the internal variables $\Omega$. The derivation turns out to
be formally simpler if, in the spirit of the slow variable representation
\cite{1996-OIT-JPB-L389}, $R$ is discretized first and the internal variables
$\Omega$ at a second stage.

The radial interval is partitioned into $M$ subintervals or elements. We generate
in each subinterval $m$ a set of $P_m$ Gauss-Lobatto points and weights
$(R_p^{(m)}, w_p^{(m)}) , p=1, \dots ,P_m$, with $R_1^{(m)}$ and $R_{P_m}^{(m)}$ 
the subinterval endpoints~\cite{BK-2001-JPB}. Note that since the endpoints of
contiguous intervals are such that 
$R_{P_{m-1}}^{(m-1)}=R_1^{(m)}$, 
the number of
{\it distinct} points in the complete grid is 
       {$L= \sum_{m=1}^M(P_m-1) + 1  $.}
The local points and weights can be used to implement the Gauss-Lobatto quadrature 
\begin{equation}
\label{quadrature}
\int_{  R_1^{(m)} } ^{ R_{P_m}^{(m)}  } f(R) dR = \sum_{p=1}^{P_m} w_p^{(m)}  f( R_p^{(m)})  ,
\end{equation}
an integration rule exact for polynomials up to degree $2P_m-3$.
Each point can be associated with a Gauss-Lobatto cardinal or shape
function defined such that $C_i^{(m)}(R_p^{(m)})=\delta_{ip}$ at the
nodal points inside the element and continued as identically zero outside
the element, $C_i^{(m)}(R)=0$ if $R \notin  [R_1^{(m)} , R_{P_m}^{(m)}]$.
The $C_i^{(m)}$ functions can be obtained by linear mapping in terms of
the corresponding cardinal functions $c_i(x)$ of the primitive interval $x \in
[-1,1]$
\begin{equation} 
\label{mapping}
C_i^{(m)}(R)=
%  c_j\left[ 2/(r_N^{(l)}-r_0^{(l)})(R -r_0^{(l)}) -1 \right] 
  c_i\left( 2~\frac{R -R_1^{(m)}}{R_{P_m}^{(m)}-R_1^{(m)}} -1 \right).
\end{equation}
% Explicit expression of the $c_i$ can be found {\it e.g.} in Ref.~\cite{BK-2001-JPB}.
Explicitly, the $c_i$ for $N$ grid points can be expressed in terms of the derivatives of
the Legendre polynomial of order $N-1$ as follows~\cite{BK-2001-JPB}
\begin{equation}
c_i(x)=\frac{-(1-x^2)}{N(N-1)P_{N-1}(x_i)(x-x_i)}\frac{dP_{N-1}(x)}{dx} ,
\end{equation}
where the Gauss-Lobatto points $x_i$ in the primitive interval comprise the endpoints $\pm 1$ and the $(N-2)$ zeros of the
$\frac{dP_{N-1}(x)}{dx}$ polynomial.

The Gauss-Lobatto cardinal functions associated to the internal points
$p=2,\cdots,P_m-1$ vanish at the element endpoints $R=R_1^{(m)} ,
R_{P_m}^{(m)}$ and following Ref.~\cite{canuto2007spectral_complex} will be referred to
as \lq\lq internal functions\rq\rq.  We also conventionally consider
as internal the cardinal functions relative to the first
$R_1^{(1)}$ and last $R_{P_M}^{(M)}$ grid points. For each internal
point, that is for $m=2,\ldots,M-1$ and $i=2,\ldots,P_m-1$ as well as
for $(i,m)=(1,1)$ and $(i,m)=(P_M,M)$, we will simply take as basis functions
the cardinal functions
\begin{equation} 
\label{basis1}
{\cal C}_i^{(m)}(R)= C_i^{(m)}(R). 
\end{equation} 
The construction of the basis functions associated to the remaining
$(M-1)$ inter-element points $R_{P_{m-1}}^{(m-1)}=R_1^{(m)}$ with
$m=2,\cdots,M$ are obtained by \lq\lq glueing\rq\rq cardinal functions
\cite{canuto2007spectral_complex}.
These interface or bridge functions are defined by
\begin{equation} 
\label{basis1a}
{\cal C}_1^{(m)}(R)=
 \left\{
        \begin{array}{ll}
  C_{P_{m-1}}^{(m-1)}(R)   &  R \in  [R_1^{(m-1)} , R_{P_{m-1}}^{(m-1)}]  \\
  C_1^{(m)}(R)   &  R \in  [ R_1^{(m)} , R_{P_m}^{(m)}]  
        \end{array}
\right.
\end{equation} 
for $m=2,\ldots,M$.
Note that contiguous subintervals are only connected through such interface functions.

In order to build a global representation of the Hamiltonian, we now introduce a
single index $a=1, \ldots, L$ running over the $L$ distinct points of
the full grid and note $R_a$ such distinct grid points. 
We define global weights
\begin{eqnarray}
\omega_a =
\left\{
	\begin{array}{ll}
		\left(w_{P_{m-1}}^{(m-1)}+w_1^{(m)}\right)  &     \text{if $R_a$ is inter-element} \\
		w_p^{(m)}  &   \text{otherwise}   .
	\end{array}
\right.
\end{eqnarray}
Similarly, we build a global grid basis comprising internal and interface functions
\begin{eqnarray}
\label{basis2}
{\cal C}_a(R) =
\left\{
	\begin{array}{ll}
		{\cal C}_1^{(m)}  &   \text{if $R_a$ is inter-element} \\
		{\cal C}_p^{(m)}  &   \text{otherwise}   .
	\end{array}
\right.
\end{eqnarray}
Equations.~(\ref{basis1}) and (\ref{basis2}) guarantee that
the orthogonality relation
\begin{equation}
\int_{  R_1 } ^{ R_L  } {\cal C}_b(R)  {\cal C}_a(R)  dR = \delta_{ba}  \omega_a
\end{equation}
holds with at least Gaussian quadrature accuracy; See
Eq.~\eqref{quadrature}.

For each value of the internal coordinates $\Omega$ we now develop the system wavefunction on the radial basis
\begin{equation}
\label{expansion}
  \Psi(R,\Omega) = \sum_{a=1}^{L} \Phi_a(\Omega) {\cal C}_a(R),
\end{equation}
the coefficients being equal to the wavefunction evaluated at the grid points $\Phi_a(\Omega)= \Psi(R_a,\Omega)$.
The Schr{\"o}dinger equation is now projected on the basis functions ${\cal C}_a$.
% $\mbox{}$\\
% $\mbox{}$\\
The second derivative arising from the radial kinetic energy term gives
rise to an integral in $R$ that is further developed as a sum of integrals
restricted to each element
\begin{eqnarray}
\label{eq_cd2psi}
\int_{R_1} ^{R_L} {\cal C}_a(R) \frac{\partial^2\Psi(R,\Omega)}{\partial R^2} dR & = & 
\sum_{m=1}^{M} \int_{R_1^{(m)}} ^{R_{P_m}^{(m)}} {\cal C}_a(R) \frac{\partial^2\Psi(R,\Omega)}{\partial R^2} dR \\
& = & \sum_{m=1}^{M} \left( 
-\int_{R_1^{(m)}} ^{R_{P_m}^{(m)}} \frac{\partial {\cal C}_a(R)}{\partial R} \frac{\partial\Psi(R,\Omega)}{\partial R} dR + \left[ {\cal C}_a(R) \frac{\partial\Psi(R,\Omega)}{\partial R}\right]_{R_1^{(m)}}^{R_{P_m}^{(m)}}
\right) , \nonumber
\end{eqnarray}
where one integration by parts has been performed for the second equality.
Noticing that for two consecutive elements one has
\begin{equation}
\label{eq_raccord}
{\cal C}_a(R_{P_m}^{(m)}) \frac{\partial\Psi(R_{P_m}^{(m)},\Omega)}{\partial R} = 
{\cal C}_a(R_1^{(m+1)}) \frac{\partial\Psi(R_1^{(m+1)},\Omega)}{\partial R} 
\end{equation}
Eq.~(\ref{eq_cd2psi}) reduces to
\begin{eqnarray}
\int_{R_1} ^{R_L} {\cal C}_a(R) \frac{\partial^2\Psi(R,\Omega)}{\partial R^2} dR   &= & 
-\sum_{m=1}^{M} \int_{R_1^{(m)}} ^{R_{P_m}^{(m)}} \frac{\partial {\cal C}_a(R)}{\partial R} \frac{\partial\Psi(R,\Omega)}{\partial R} dR  \\
 & & + {\cal C}_a(R_{L}) \frac{\partial\Psi(R_L,\Omega)}{\partial R}
- {\cal C}_a(R_{1}) \frac{\partial\Psi(R_1,\Omega)}{\partial R} . \nonumber
\end{eqnarray}
Note that the boundary terms cancellation of Eq.~\eqref{eq_raccord}
holds for the exact solution but is only approximatly valid when the
solution is computed as an expansion on a finite basis. In other terms,
the numerical solution will in general have a discontinous derivative
at the element interfaces. However, such left-right discontinuity tends
to zero exponentially for a converged calculation and as such does not
affect the fast convergence rate demonstrated in Sec.~\ref{numerical_tests}.
Using the decomposition Eq.~\eqref{expansion} for the evaluation of
$\partial \Psi(R,\Omega)/\partial R$ one gets a term involving the matrix
\begin{equation}
\label{eq_mattperelement}
{\cal T}_{ab}  = \sum_{m=1}^{M} 
\int_{  R_1^{(m)} } ^{  R_{P_m}^{(m)}  }
\frac{d{\cal C}_a(R)}{dR}  \frac{d {\cal C}_b(R)}{dR} \,dR  
\end{equation}
formally recast as
\begin{equation}
\label{eq_matt}
{\cal T}_{ab}  = 
\int_{  R_1 } ^{  R_L  }
\frac{d{\cal C}_a(R)}{dR}  \frac{d {\cal C}_b(R)}{dR} \,dR  .
\end{equation}
Using the definitions Eq.~(\ref{basis1}) and Eq.~(\ref{basis1a}), the linear mapping in Eq.~(\ref{mapping}),
and approximating the integrals on the rhs by the quadrature
of Eq.~\eqref{quadrature}, the kinetic matrix can be expressed in terms of
analytically  known Gauss-Lobatto derivation matrices $C_j^\prime(x_i)
$; See {\it e.g.} \cite{BK-2001-JPB}.

A closer look at Eq.~(\ref{eq_mattperelement}) keeping into account
the local character of the ${\cal C}_a(R)$ functions shows that most
elements of ${\cal T}$ are zero. More specifically, ${\cal T}_{ab}=0$
if ${\cal C}_a$ and ${\cal C}_b$ are both internal functions and do not
belong to the same element. If ${\cal C}_a$ is an interface function,
thus at the interface of two elements $m$ and $m+1$, ${\cal T}_{ab}=0$ if
$b$ does not belong to any of the two $m$ and $m+1$ elements, while ${\cal
T}_{ab}\neq0$ if ${\cal C}_b$ is an internal or an interface function
of element $m$ or  $m+1$.  In addition, ${\cal T}_{ab}\neq0$ if both
$a$ and $b$ belong to the same element. 
The potential energy is approximately diagonal in the grid basis with diagonal elements
$\omega_a { V}(R_a, \Omega)$ if quadrature Eq.~\eqref{quadrature} is used.
Collecting all the terms, the matrix form of the Schr\"odinger equation finally reads
\begin{equation}
\label{schroedinger1}
\sum_{b=1}^{L} {\cal T}_{ab} \Phi_b(\Omega) + \frac{2 \mu}{\hbar^2} \omega_a \left[ { V}(R_a, \Omega)  -E \right]  \Phi_a(\Omega)
=  {\delta_{La}} {\partial_R \Psi}|_{R=R_L} - {\delta_{1a}} {\partial_R \Psi} |_{R=R_1}.
\end{equation}

We now introduce an internal coordinate basis $\phi_{\alpha}(\Omega)$
whose nature or dimension does not need for the moment to be
specified. 
We write therefore
\begin{equation}
\label{expangle}
  \Phi_a(\Omega) = \sum_{\alpha=1}^{N^{(a)}}  F_{a \alpha} \phi_{\alpha}^{(a)}(\Omega) ,
\end{equation}
where the superscript $(a)$ stresses the possible dependence of the basis on the grid point.
Insertion of Eq.~\eqref{expangle} in Eq.~\eqref{schroedinger1} leads to the algebraic equations
\begin{equation}
\label{eq_fullmat}
 \sum_{b=1}^L  \sum_{\beta=1}^{N^{(b)}}  {\cal T}_{ab}   {\cal O}_{a \alpha, b \beta} \,  F_{b \beta }  + \frac{2\mu}{\hbar^2} \sum_{\beta=1}^{N^{(a)}} \omega_a \left[  {\cal U}_{\alpha 
 \beta}(R_a)  -E \delta_{\alpha \beta} \right] F_{a \beta }  
 = {\delta_{La}}   \xi_{\alpha L} -  {\delta_{1a}}  \xi_{\alpha 1} 
\end{equation}
where ${\cal O}_{a \alpha, b \beta}=\langle \phi_{\alpha}^{(a)} |
\phi_{\beta}^{(b)} \rangle_\Omega $ is the overlap matrix element
over the $\Omega$ coordinates and ${\cal U}_{\alpha \beta}(R_a) =
\langle \Phi_\alpha^{(a)} (\Omega)| V(R_a,\Omega) | \Phi_\beta^{(a)}
(\Omega)\rangle $. The quantities $\xi_{\alpha a}$ with $a=1,L$ are the
normal derivatives of the wavefunction at the integration boundaries in
channel $\alpha$, {\it i.e.}
\begin{equation}
    \xi_{a \alpha } = \langle \phi_{\alpha}^{(a)}  |   {\partial_R \Psi}|_{R=R_{a}}  \rangle_\Omega .%\quad  \text{  with  } \quad (a,l)=(1,1),(L,M).
\end{equation}

Equations~\eqref{eq_fullmat}, supplemented by the proper boundary conditions
in Sec.~\ref{sec_bs} and~\ref{sec_ss}, represent the key formal result of the paper.
In order to maximize sparsity one can require that the
basis $\phi_\alpha^{(a)}$ does not depend on the grid point $a$.  This is
for example the case when using spherical or hyperspherical harmonics,
or ro-vibrational molecular states. In this case, the ${\cal O}$ matrix
reduces to the identity matrix. 

\begin{figure}[!hbt]
\centerline{\epsfig{file=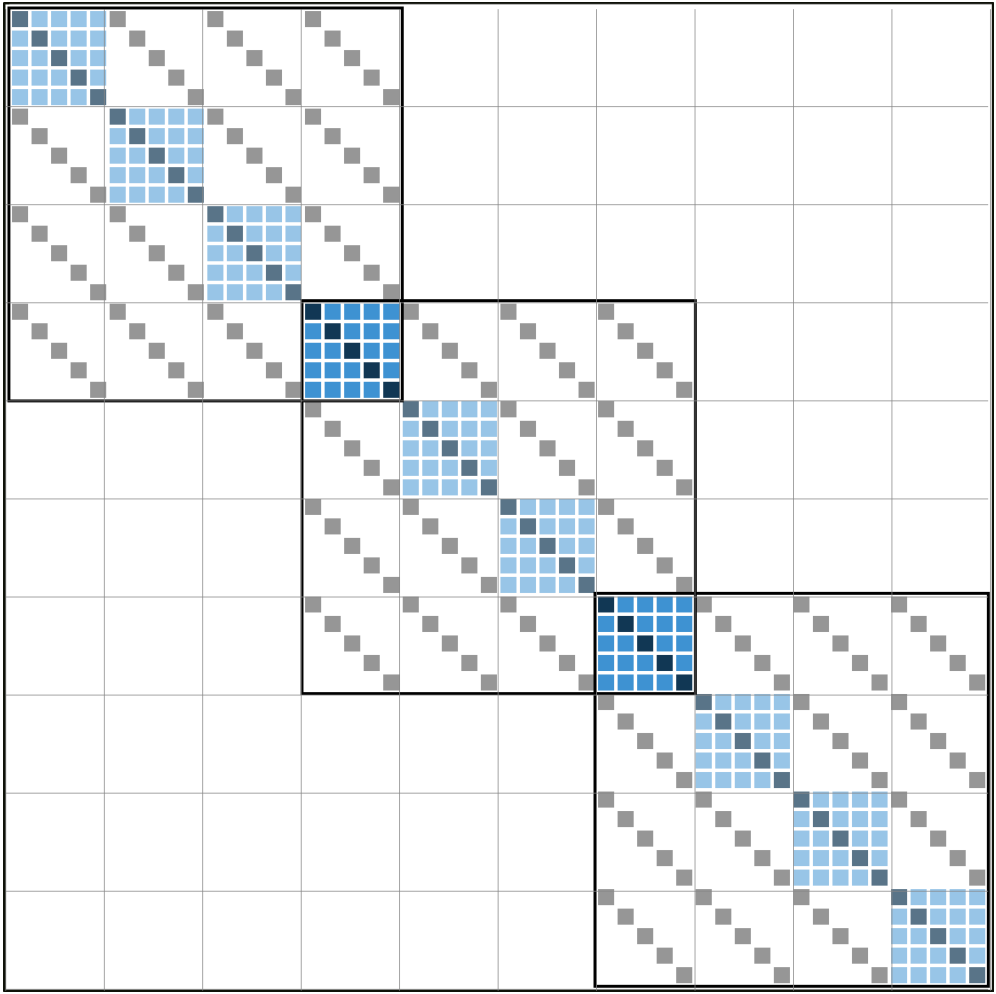,width=.80\columnwidth,angle=0}}
\caption[]{
(Color online) Structure of the discretized matrix for 3 elements, 4
Lobatto points and 5 channels per element. Each full block along the
main diagonal (blue) contains the 25 channel matrix elements ${\cal
T}_{aa}   {\delta}_{\alpha \beta}  + \frac{2\mu}{\hbar^2}  \omega_a
\left[  {\cal U}_{\alpha
 \beta}(R_a)  -E \delta_{\alpha \beta} \right]$ for $\alpha,\beta=1,\dots,
 5$ on the lhs of the equation system Eq.~\eqref{eq_fullmat}. Each off-diagonal
diagonal block (grey) only arises from kinetic energy coupling and
contains matrix elements $ {\cal T}_{ab}   {\delta}_{\alpha \beta} $.
The two 5x5 matrix blocks in darker color contained in the main diagonal indicate
inter-element points connecting contiguous elements (see text). 
}
\label{fig_matrix_simple}
\end{figure}

A pictorial representation of the resulting Hamiltonian matrix is given
in Fig.~\ref{fig_matrix_simple} where each small block correspond to
fixed grid indices and varying channel indices. Similar matrix structure
representations can be found elsewhere in the literature, for instance
in~\cite{2010-AS-PRA-053845,2010-SXH-PRA-056705}.
For this example, three subintervals $M=3$ are considered,
the number of Gauss Lobatto points and
the number of basis for the internal coordinates are identical for the
three subelements and are fixed to $P_m=4$ and $N^{(a)}=5$ channels.
The non-zero elements arising from the kinetic part are depicted in
gray while the ones resulting from the potential are in blue. For
the illustrative case presented, 340 over the 50x50 matrix elements
are non-zero, which amounts to a filling factor of 13.6\%. 

Restoring some flexibility in the choice of the basis used for the
$\Omega$ part but still leaving quite a large sparsity in the full
Hamiltonian matrix can also be obtained by imposing that the basis
does not vary within each element. This means that the dependence of
$\phi_{\alpha}^{(a)}$ on $(a)$ is replaced by an ensemble of basis
functions in which the same basis is used for all points within
the same element $m$ with the exception of the inter-element points
for which alternative basis functions may be used. One such example
would be the set of eigenvectors obtained through diagonalization of
the reduced Hamiltonian ${ V}(R_{x},\Omega)$ at a fixed point $R_x$
inside the element (diabatic-by-sector method~\cite{1989-JML-CPL-178}).
The diabatic-by-sector approach trades loss of some sparsity with a
(possible) reduction in the basis size.

\begin{figure}[!hbt]
\centerline{\epsfig{file=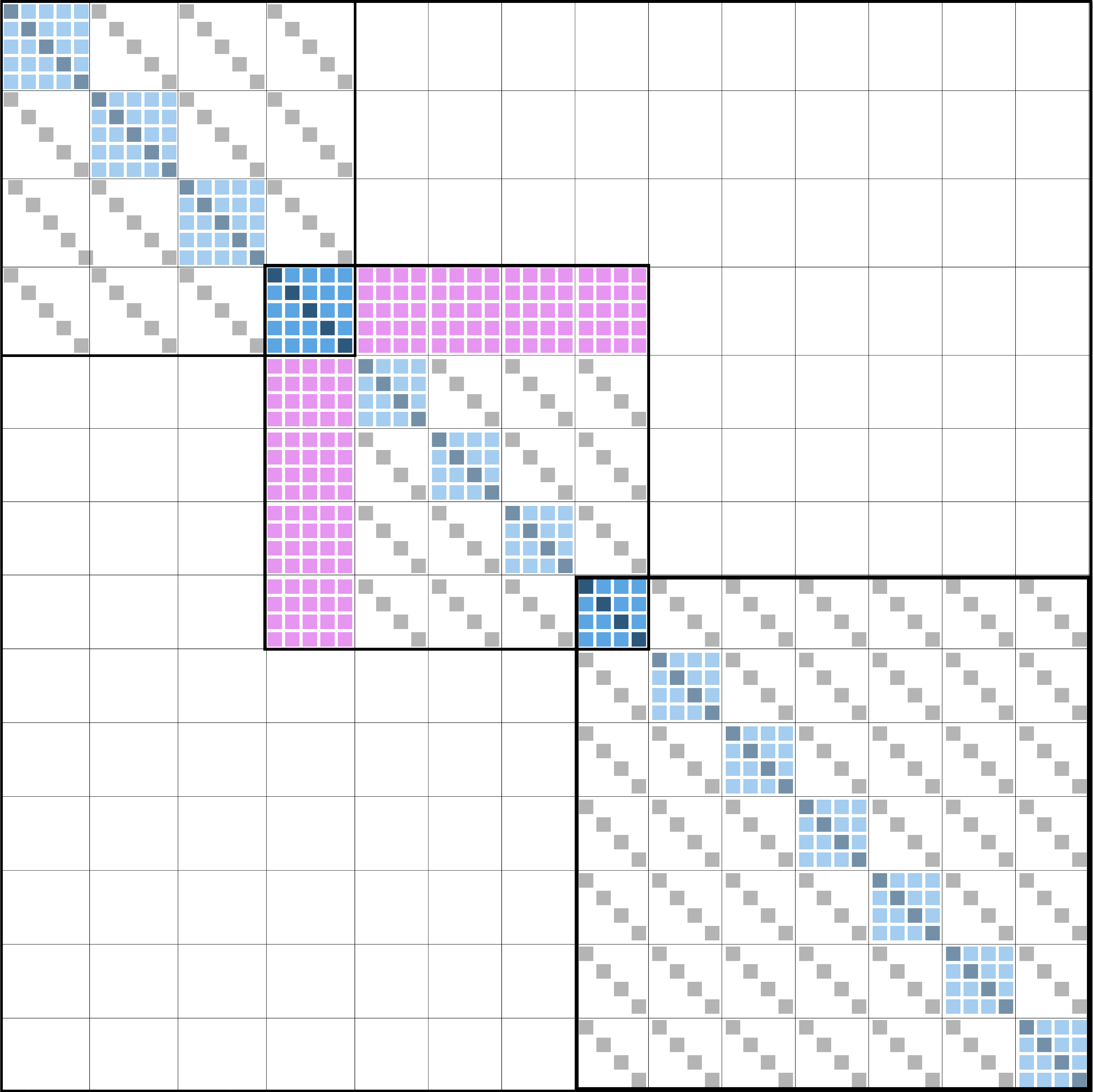,width=.80\columnwidth,angle=-90}}
\caption[]{(Color online)
Same as in Fig.~\ref{fig_matrix_simple} with 3 elements but now varying
number of Lobatto points in each element (4, 5, and 6, respectively)
and different channel bases in elements 1 and 2.  The overlap matrix
in Eq.~\eqref{eq_fullmat} is now a full matrix giving rise to the full
off-diagonal blocks (violet) containing elements ${\cal T}_{ab}   {\cal
O}_{a \alpha, b \beta}$.
}
\label{fig_matrix_complex}
\end{figure}

As an example, the matrix structure for a more flexible basis choice 
is depicted in Fig.~\ref{fig_matrix_complex} in which different
numbers of Lobatto points are used $P_1=4$, $P_2=5$ and $P_3=7$. In
%addition the first element contains 5 channels whereas the other two
%4 alternative channels. The channel basis associated to the first
addition the first element contains 5 basis functions whereas the other two
4 alternative basis functions. The  basis associated to the first
%inter-element point is taken to be identical to the 5 channels of the
%first element. The modification of the channel basis between elements 1
inter-element point is taken to be identical to the 5 basis functions of the
first element. The modification of the  basis between elements 1
and 2 induces additional non-zero elements (pink in the figure) due to
basis overlaps.

In the extreme case where the Hamiltonian is diagonalized at each point,
one retrieves the slow-variable formulation proposed for hyperspherical
bound states in Ref.~\cite{2011-HS-JCP-064318}. Such fully adiabatic
procedure does optimize the basis size, but results in full overlap
matrices at {\it all} off-diagonal grid elements, putting severe memory
constraints on the size of treatable systems.


Realistic calculations usually require large number of elements and
points to be converged. Therefore, in the locally diabatic formulation
the empty part of the matrix becomes large and the filling factor
decreases significantly. More quantitatively, let us consider a potential
represented by a full matrix with $N$ channels. For a partition
composed of $M$ elements with $P$ points per element the number of
nonzero elements is $N(N+1) \left[ M(P-1)+1 \right] /2 + P(P-1)MN/2
$, that reduces to $\approx LN\left[ N+P+1  \right]/2 $ for $M,P \gg
1$. Note that due to matrix symmetry only elements above (or below)
the main diagonal have been taken into account. The total number of
elements (now both above and below the main diagonal) is $N^2 \left[ M(P-1)
+1  \right]^2 \approx (NLP)^2$, resulting in a filling factor $\approx
(N+P+1)/(2 NLP^2)$ that may easily drop below 1\%. We remark that
this worst case scenario of a full potential matrix seldom happens
in molecular physics due to the tensor nature of at least part of the
interaction and to the accompanying selection rules.

Depending on the problem, purely diabatic and diabatic-by-sector
representation can also be conveniently combined. For instance, in
Ref.~\cite{2015-AS-NJP-013020} the present algorithm was used to join
a purely diabatic spherical harmonics basis at short range with a
diabatic-by-sector representation at long range to treat the difficult
problem of ultracold collisions between two polar molecules in an
optical tube.

\subsection{Bound states}
\label{sec_bs}

Bound states $\Psi^{(n)}$ of the system with energy $E_n$ are calculated
imposing that the solution of the Schr\"odinger equation vanishes at
the boundaries $R_1$ and $R_L$ of the radial interval. Accordingly,
for $a=1$ and $L$ the expansion coefficients $F_{a \alpha}^{(n)}=0$, and
all elements of the discretized Hamiltonian with grid indices $a,b=1,L$
in the system of equations \eqref{eq_fullmat} can be dropped.

The resulting equations for the multichannel bound-state solution 
at the remaining $L-2$ points present themselves in the form of
a generalized eigenvalue problem
\begin{equation}
\label{eq_bs}
 \sum_{b=2}^{L-1}  \sum_{\beta=1}^{N^{(b)}}  {\cal T}_{ab}   {\cal
 O}_{a \alpha, b \beta} \,  F_{b \beta }^{(n)}  + \frac{2\mu}{\hbar^2}
 \sum_{\beta=1}^{N^{(a)}}  \omega_a {\cal U}_{\alpha \beta}(R_a)  F_{a
 \beta }^{(n)} =   E_n  \omega_a F_{a \alpha }^{(n)}  \quad , \quad   a=2,\dots,L-1  .
\end{equation}
The $\omega_a$ factor on the rhs can be removed by redefining as new unknown
$\sqrt{\omega_a} F_{a \alpha }^{(n)}$ and by right multiplication by the
diagonal matrix with element $1/\sqrt{\omega_a}$. In this way, the problem
is expressed as an ordinary eigenvalue problem.
Finally note that, if needed, equations \eqref{eq_fullmat} with left grid
index $a=1$ and $L$ can be used to compute the normal derivative components
$\xi_{a \alpha}^{(n)}$ as a function of the $F_{b \alpha }^{(n)}$
with $b=2,\dots,L-1$.

\subsection{Scattering states}
\label{sec_ss}

In the case of scattering solutions, a number $N^{(L)}$ of
linearly independent solutions with energy $E$ can by built from
Eq.~\eqref{eq_fullmat}. We consider for definiteness the most common
case where at the left end point $R_1$ the wavefunction vanishes. As
in the bound-state problem, this implies that all lines and columns in
the system of equations \eqref{eq_fullmat} with grid index $a=1$ can be
dropped. At the other edge of the grid, we discuss below two approaches
to obtain the physical wavefunction and thus the relevant scattering
observables. The first one requires to compute a matrix comprising all
linearly independent solutions of the discretized Schr\"odinger equation,
the second one can be used to determine a single column-vector solution.

\subsubsection{$R$-matrix boundary conditions}

The so-called $R$-matrix solutions ${\bar \Psi}^{(\gamma)}$ are defined by the
condition that their normal derivative vanishes on the surface $R=R_L$
in all but channel $\gamma$, where it is unity. Therefore, such $N^{(L)}$
independent $R$-matrix solutions with energy $E$ can be determined by
imposing $ \xi_{\alpha L} = \delta_{\alpha \gamma}$ and solving the
linear system
\begin{equation}
\label{eq_rmat}
 \sum_{b=2}^L  \sum_{\beta=1}^{N^{(b)}}  {\cal T}_{ab}   {\cal O}_{a \alpha, b \beta} \,  {\bar F}_{b \beta }^{(\gamma)}  + \frac{2\mu}{\hbar^2} \sum_{\beta=1}^{N^{(a)}} \omega_a \left[  {\cal U}_{\alpha 
 \beta}(R_a)  -E \delta_{\alpha \beta} \right] {\bar F}_{a \beta }^{(\gamma)}  
 =    {\delta_{La}}  \delta_{\alpha \gamma}  \quad , \quad a=2,\dots,L  .
\end{equation}
The $R$-matrix $\mathbf R$ is simply defined as the matrix with elements
$R_{\alpha \beta} = {\bar F}_{L \alpha }^{(\beta)}$.

Solutions with physical boundary conditions can be written for $R \geq
R_L$ as a superposition of solutions of the asymptotic Hamiltonian,
comprising angular channel eigenfunctions $\Phi_{\alpha}^{(L+1)}$ and
of reference regular $\tilde f$ and irregular $\tilde g$ radial functions
\begin{equation}
\label{eq_asy}
\Psi^{(I)}(R, \Omega)= \sum_{\alpha=1}^{N^{(L)}} \left[ {\tilde f}_{\alpha}(R) \delta_{\alpha I}  
                           -  {\tilde g}_{\alpha}(R) K_{\alpha I} \right ] \Phi_{\alpha}^{(L+1)}(\Omega) .
\end{equation}
The channel eigenfunctions are $R$-independent and the superscript $(L+1)$
is merely introduced as an additional artificial grid point for ease of
notation in subsequent formal manipulations. A solution $\Psi^{(I)}$
corresponds to a wave incoming in channel $I$ with scattered
waves in all channels $\alpha$, with amplitudes $K_{\alpha I}$. If $f$
and $g$ are real standing waves the coefficients $K_{\alpha I}$ form the
reaction matrix $\mathbf K$. 

The solutions $\Psi^{(I)}$ and its normal derivative can be expressed
on the surface $R=R_L$ as linear combinations of the $R$-matrix solutions
${\bar \Psi}^{(\gamma)}$ with constant coefficients $N_{\gamma I}$
\begin{equation}
\label{eq_match1}
\Psi^{(I)}(R_L, \Omega) = \sum_{\gamma=1}^{N^{(L)}}  {\bar \Psi}^{(\gamma)}(R_L, \Omega)  N_{\gamma I}
\end{equation}
and 
\begin{equation}
\label{eq_match2}
\partial_R \Psi^{(I)}(R_L, \Omega) = \sum_{\gamma=1}^{N^{(L)}} \partial_R  {\bar \Psi}^{(\gamma)}(R_L, \Omega)  N_{\gamma I} .
\end{equation}

Following the standard asymptotic matching procedure \cite{1989-JML-CPL-178}, the
expression Eq.~\eqref{eq_asy} is inserted on the lhs of Eqs.~\eqref{eq_match1}
and \eqref{eq_match2} and the latter are projected on the angular basis
$\Phi^{N^{(L)}}(\Omega)$. The resulting linear system can be easily
solved for $\mathbf K$ in terms of $\mathbf R$
\begin{equation}
\label{eq_asysolve}
  \mathbf K = \left( {\mathbf g} - {\mathbf R} {\mathbf g}^\prime \right)^{-1}  \left( {\mathbf f} - {\mathbf R} {\mathbf f}^\prime \right)  .
\end{equation}
Here, matrices $\mathbf f$ and ${\mathbf f}^\prime$ are respectively defined as
\begin{equation}
 f_{\alpha \beta}={\tilde f}_\alpha(R_L) {\cal O}_{L \alpha, (L+1)\beta}   \quad , \quad f_{\alpha \beta}^\prime= {\tilde f}_\alpha^\prime(R_L) {\cal O}_{L \alpha, (L+1)\beta}
\end{equation}
as a function of the overlap between the asymptotic channels and the angular
basis at last grid point. A similar definition holds for $\mathbf g$ and ${\mathbf g}^\prime$.

\subsubsection{Scattering boundary conditions}

Rather than going through the determination of $N^{(L)}$ independent $R$-matrix solutions, scattering 
boundary conditions can also be incorporated directly in the linear system of equation \eqref{eq_fullmat}.
To this aim, we first impose that at last grid point $a=L$ the wavefunction takes the form~\eqref{eq_asy}
\begin{equation}
 F_{L \alpha}^{(I)} = \sum_{\beta}\left[  f_{\alpha \beta}  \delta_{\beta I} - g_{\alpha \beta}  K_{\beta I} \right] .
\end{equation}
Similarly, the normal derivative channel components on the rhs of Eq.~\eqref{eq_fullmat} becomes
\begin{equation}
\label{eq_xi}
 \xi_{L \alpha}^{(I)} = \sum_{\beta}\left[  f^\prime_{\alpha \beta}  \delta_{\beta I} - g^\prime_{\alpha \beta}  K_{\beta I} \right]  .
\end{equation}

For notational ease, we define the matrix on the lhs of Eq.~\eqref{eq_fullmat} 
\begin{equation}
M_{a \alpha,b \beta} = {\cal T}_{ab}   {\cal O}_{a \alpha, b \beta} + \frac{2\mu}{\hbar^2} \omega_a \left[  {\cal U}_{\alpha \beta}(R_a)  -E \delta_{\alpha \beta} \right] \delta_{ab}
\end{equation}

As it will be clear from equation \eqref{eq_smat} below, in order to
obtain a symmetric linear system, it is necessary to introduce the new
unknown ${\mathbf X}= {\mathbf g } {\mathbf K}$ in the place of $\mathbf
K$. We also define the log-derivative ratio ${\mathbf Y}^g= {\mathbf g
}^\prime {\mathbf g }^{-1}$, such that the quantity ${\mathbf g }^\prime
{\mathbf K}$ on the rhs of Eq.~\eqref{eq_xi} becomes ${\mathbf g }^\prime
{\mathbf K} = {\mathbf Y}^g {\mathbf X}$.

With these definitions, simple matrix algebra allows one to cast the system~\eqref{eq_fullmat} into the form
\begin{equation}
\label{eq_smat}
\sum_{b=2}^{L-1} \sum_{\beta=1}^{N^{(b)}} M_{a \alpha,b \beta} F_{b
\beta}^{(I)} + \sum_{\beta=1}^{N^{(L)}} \left[  M_{a \alpha,L \beta}
+ {\delta_{La}} Y_{\alpha \beta}^g \right]
X_\beta^{(I)} =  - \sum_{\beta=1}^{N^{(L)}}  M_{a \alpha,L \beta} f_{\beta
I} + {\delta_{La}} f_{\alpha I}^\prime.
\end{equation}

As a final step, the $K$-matrix can be computed from the definition
of ${\mathbf X}$ by solving the linear system ${\mathbf g } {\mathbf
K}={\mathbf X}$. It is important to notice that for a given incoming
wave labeled by index $I$ one can determine a single column of the
matrix solution ${\mathbf X}$ and thus of $\mathbf K$. If one uses
complex algebra and replaces $f_\alpha$ and $g_\alpha$ by travelling waves $h_\alpha^{(-)}$
and $h_\alpha^{(+)}$, the asymptotic condition~\eqref{eq_asy} becomes
\begin{equation}
\Psi^{(I)}(R, \Omega)= \sum_{\alpha=1}^{N^{(L)}} \left[ h_{\alpha}^{(-)}(R) \delta_{\alpha I}
                           -  h_{\alpha}^{(+)}(R) S_{\alpha I} \right ] \Phi_{\alpha}^{(L+1)}(\Omega) .
\end{equation}
with $\mathbf S$ the scattering matrix, whose elements are directly
related to observables. 
The procedure to determine $\mathbf K$ presented in this section applies
as is to the determination of $\mathbf S$, leading to the equivalent of
Eq.~\eqref{eq_smat} with ${\mathbf f}$ and ${\mathbf g}$ replaced by ${\mathbf h^{(-)}}$ and ${\mathbf h^{(+)}}$
and ${\mathbf Y}^g$ by ${\mathbf Y}^{h^{(+)}}= {\mathbf h}^{(+) \prime}
 [{\mathbf h }^{(+)}]^{-1}$.
Determining a single column of interest of the scattering matrix may
lead to computational advantages, in particular in problems with large
numbers of open channels.

\subsection{Spectral log-derivative propagation}

In spite of the sparse character of the discretized Hamiltonian, memory
can become a limiting factor for systems described by large numbers of
collision channels. In this case, it may be necessary to split the full
propagation interval in smaller intervals, each comprising for instance
only one element. The scattering equation is solved in any given element
element to determine at each point a matrix of linearly independent
solutions ${\mathbf F}_a$ with elements $F^{(I)}_{\alpha a}$ labeled
by column index $I$ and channel index $\alpha$. Such solutions will
be combined to form the log-derivative matrix ${\mathbf Y}_a={\mathbf
F}_a^{\prime} {\mathbf F}_a^{-1}$.

Our main equation \eqref{eq_fullmat} specialized to an element with
$P_m$ points can now be rearranged as an algorithm expressing the value
of ${\mathbf Y}_{P_m}$ on the right-end of the element to a known input
value ${\mathbf Y}_1$ assigned on the left-end point. This task can
be accomplished by right multiplications by ${\mathbf F}_{P_m}^{-1} $
to give after simple algebra :

\begin{equation}
\label{eq_logder_a}
\sum_{b=1}^{P_m-1} \sum_{\beta=1}^{N^{(b)}} \left[  M_{a \alpha,b \beta} + {\delta_{1a}} Y_{1,\alpha \beta} \right]    {\bar F}_{b,
\beta \gamma} = - M_{a \alpha,P_m \gamma}
\quad , \quad  a=1,\dots,(P_m-1)   .
\end{equation}
With $\bar{\mathbf F}_a \equiv {\mathbf F}_a {\mathbf F}_{P_m}^{-1}$ determined at first $(P_m-1)$ points, the remaining equation
at last point
\begin{equation}
\label{eq_logder_b}
 Y_{P_m,\alpha \gamma}  =  \sum_{b=1}^{P_m-1}
\sum_{\beta=1}^{N^{(b)}} M_{P_m \alpha,b \beta} {\bar F}_{b, \beta \gamma}
 +M_{P_m \alpha,P_m \gamma} 
\end{equation}
determines the final log-derivative through a series of sparse matrix multiplications.
The log-derivative ${\mathbf Y}_{P_m}$ can then be used as entry for the
calculation in next element.

\subsection{Error control}

A major advantage of the spectral element method is the possibility
to estimate precisely the numerical error by {\it a posteriori}
analysis of the calculated solution. To this aim, let us consider
the solution wavefunction restricted to the $m$-th sector and suppose
for the sake of simplicity that the angular basis of dimension $N_m$
is constant within the sector. The discretized solution at the $P_m$
points in the sector for the different channels is therefore represented
by $P_m \times N_m$ elements noted $F_{p \alpha}$.

The error is estimated by first performing an orthogonal
transformation from the grid basis to the polynomial basis
of Legendre polynomials $P_n(R)$ defined in the $[ R_1^{(m)},
R_{P_m}^{(m)} ]$ interval through a coordinate transformation
in the same fashion as in Eq.~\eqref{mapping}. Assuming the
polynomials normalized, the transformation matrix reads explicitly
$O_{np}=P_n(R_p)/\sqrt{w_p^{(m)}}$. For each channel $\alpha$
the transformation $\mathbf O$ gives the set of pseudospectral
coefficients ${\tilde F}_{n \alpha}$ of the solution expanded on the
Legendre basis as ${\tilde F}_{n \alpha} = \sum_{n=1}^{P_m} O_{np}
F_{p \alpha}$.  The main point is that the convergence of the Legendre
polynomial series is superalgebraic, at least for sufficiently regular
solutions~\cite{BK-2001-JPB}. The size of last calculated coefficients
${\tilde F}_{P_m \alpha}$ is therefore a reliable estimate of the
remainder of the series, {\it i.e.} of the numerical truncation error
in each channel. If the error is larger (smaller) that a given tolerance
criterion one can either reduce (increase) the element size or increase
(reduce) the polynomial order $P_m$. As recognized at the birth of the
so-called $hp$-methods, the optimal strategy to guarantee an exponential
accuracy of the calculated solution consists in increasing $P_m$ in the
regions where the latter is regular and in decreasing the element size in
the regions where it is irregular~\cite{1992-IB-AES-159}.

We will show in the next section a series of numerical experiments
for both scattering and bound state calculations. We limit ourselves
to a relatively simple ro-vibrational model with a purely diabatic
basis in order to make the numerical convergence analysis as plane
as possible. Since as most usual in molecular physics the solution is
regular we fix the same polynomial order in all elements and study the
behavior of selected observables as a function of both the element size
and polynomial order.

\section{Numerical tests}
\label{numerical_tests}
We perform numerical tests of efficiency and accuracy of the
algorithm on the Rb$_2$He trimer, a system for which bound states and
ultra-cold scattering properties have already been studied in our
group~\cite{guillon2012,viel2014}.

For the description of Rb$_2$He, the $\vec{R}$ and $\vec{r}$ Jacobi vectors are used.
The corresponding Hamiltonian in the space fixed frame reads~\cite{arthurs1960}
\begin{equation}
\hat H = - \frac{\hbar^2}{2\mu_{\text{Rb$_2$--He}}}\left( \frac{1}{R}\frac{\partial^2}{\partial R^2} R\right)
 -  \frac{\hbar^2}{2\mu_{\text{Rb$_2$}}} \left( \frac{1}{r}\frac{\partial^2}{\partial r^2} r\right)
+ \frac{ L^2}{2\mu_{\text{Rb$_2$--He}} R^2}
+ \frac{ j^2}{2\mu_{\text{Rb$_2$}} r^2}
+ \hat V,
\label{eq_H_rb2he}
\end{equation}
with $R$  and $r$ the Rb$_2$ -- He and Rb$_2$ distances,
$\mu_{\text{Rb$_2$--He}}$ and $\mu_{\text{Rb$_2$}}$  the associated
reduced masses, ${\vec{L}}$ and ${\vec{j}}$
the angular momenta and $\hat V$ the potential term taken from
Ref.~\onlinecite{viel2014} limited to the 2-body part.  The generic
$\Omega$ coordinates introduced in Sec.~\ref{sec_discre} correspond
for this system to five spatial coordinates, namely, $r$, $\hat r$
and $\hat R$ that reduce to three when fixing the total angular
momentum quantum numbers $J$ and $M$.  The basis functions used to
represent $\Phi_a(\Omega)$ in Eq.~\eqref{expangle} are taken identical for
all sectors. Equation~(\ref{expangle}) reads for this specific case
\begin{equation}
%u_a(\Omega) = \sum_{vj\ell} v_{a,vj\ell} \frac{1}{r} \chi_{vj}(r)Y_{j\ell}^{JM}\left(\hat r, \hat R \right),
\Phi_a(\Omega) = \sum_{vj\ell} v_{a,vj\ell} \frac{1}{r} \chi_{vj}(r)Y_{j\ell}^{JM}\left(\hat r, \hat R \right),
\label{eq_basis_rb2he}
\end{equation}
where $\chi_{vj}(r)$ are the rovibrational eigenstates of the Rb$_2$
diatomic and $Y_{j\ell}^{JM}$ the coupled spherical harmonics
\cite{arthurs1960}.
\subsection{Scattering states}
For the calculations, we vary $R$ from 4 to 120 $a_0$, we use
Rb$_2(v=1,j=0)$ as the initial state for the collision, and we impose
$R$-matrix boundary conditions. The linear system Eq.~(\ref{eq_rmat})
is solved using the PARDISO package~\cite{pardiso-6.0a,pardiso-6.0b}
included in the MKL library~\cite{MKL}. This state-of-the art direct
solver determines the solution of a sparse linear system in (number
of nonzero elements)$^{3/2}$ operations. In our tests, we
find that total memory used by PARDISO is about five times larger than
the memory required to store the nonzero elements of the discretized
Hamiltonian. The $K$-matrix, extracted from the matching procedure in
Eq.~\eqref{eq_asysolve} performed at $R=120$~$a_0$, is diagonalized to
compute the eigenphasesum
\begin{equation}
\delta = \sum_{i=1}^{n_{op}} \arctan (\eta_i),
\label{eq_esumphase}
\end{equation}
with $\eta_i$ being the $n_{op}$ eigenvalues of the open-open part of the
$K$-matrix. Note that while the eigenphasesum is a function of collision
energy and depends on the partial wave $J$ considered, the corresponding
indices have been dropped for ease of notation.

For the accuracy tests, we used a collisional energy of 1 K above
the $v=1,j=0$ initial state and focussed on the $J=1$ partial wave.
Basis functions with up to \text{$v=4, j=24$} quantum numbers are
included, resulting in 125 channels, 32 of which are energetically open at the
considered collision energy. A fixed Lobatto order $P$ is used for
all $M$ elements used for the discretization of the $[R_1,R_L]
= [4,120]$ (in $a_0$) interval. All elements are taken of the same
length, noted $h$ hereafter, with 
\begin{equation} 
h=(R_L-R_1) / M.
\end{equation}

\begin{figure}[!hbt]
\centerline{\epsfig{file=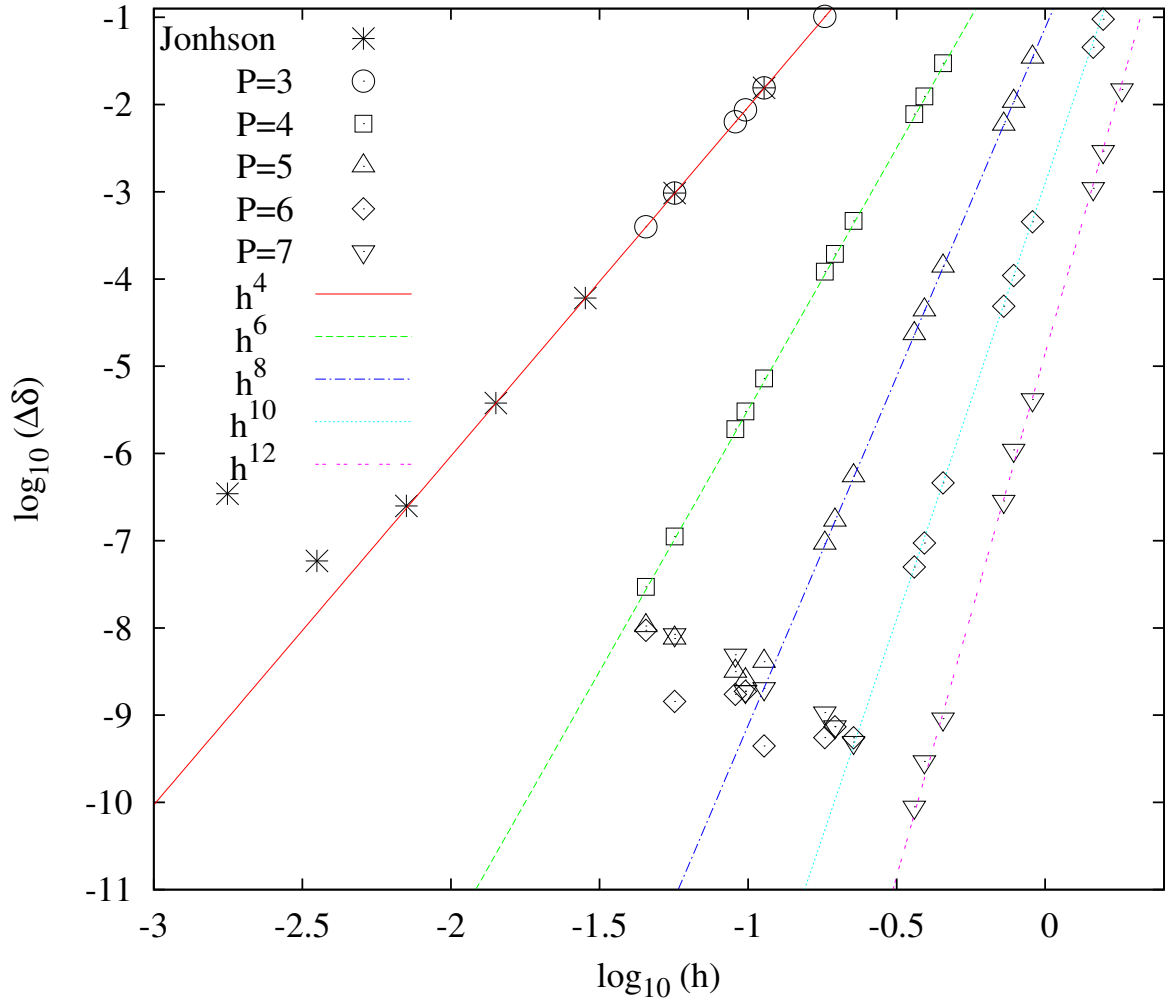,width=.99\columnwidth,angle=0}}
\caption[]{Error as a function of element size for Johnson (star) and various Lobatto order (P=3 to 7).
Relevant lines are also indicated.  }
\label{fig_error}
\end{figure}

Figure~\ref{fig_error} presents the variation of $\delta$ as a function of $h$ in log-log scale when $h$ is systematically divided by two.
Assuming a dependence of $\delta$ of the form
\begin{equation}
\label{eq_error_h}
\delta = \delta_0 + C h^\alpha,
\end{equation}
the power $\alpha$ is directly accessible by the slope of the variation
of the quantity $\Delta\delta = \delta(h)-\delta(h/2)$ as a function
of $h$ in log-log scale, even when the exact value of $\delta_0$ is unknown.
A comparison with the resolution of the coupled equations performed by the
Johnson log-derivative propagator~\cite{1973-BRJ-JCoP-445} is
also presented in the figure. The results for the Lobatto order $P=3$ case
are strictly identical to the ones obtained with the Johnson
propagator method. 
One can indeed show formally that solution of Eq.~\eqref{eq_logder_a} with
two Lobatto points followed by application of~\eqref{eq_logder_b} gives
exactly the same result as the three-points Johnson recursion. However,
the two algorithms should not be viewed as equivalent, in the sense that
the first half propagation step in Johnson's method is not equivalent
to solving our equation~\eqref{eq_logder_a} for two points.

For these cases, the known $\alpha=4$ value can be
read from the linear curves in the figure.  For $\log_{10}(h) < -2.2$ the
accuracy on $\delta$ can no longer be improved by a reduction of $h$ when
using the Johnson propagator.  For $P=3$ Lobatto case, the
memory requirement prohibits the computation at such small $h$.  For the
$P=4$ to $P=7$ Lobatto orders, a slope $\alpha=2P-2$ is obtained.
When increasing $P$, lower values of the absolute error are obtained for
specific values of $h$.  For example a $10^{-10}$ accuracy is reached for $P=7$
and $h\sim 0.4$~$a_0$. We stress that $\delta$ is a very sensitive
quantity and such absolute error value on $\delta$ corresponds to the
usually observed accuracy on rate coefficients calculations, much easier to converge.
%For larger Lobatto order, we assumed a known numerical value of $\delta_0$ and present in fig.~\ref{fig_ph} the ($h$,$N$) combinations for which the relative error on $\delta$ is lower than $1\%$.
%\begin{figure}[!hbt]
%\centerline{\epsfig{file=Fig_compute_p_h.eps,width=.99\columnwidth,angle=0}}
%\caption[]{Dots materialize relative errors on $\delta$ better than 1\% as a function of $h$ in $a_0$ and of Lobatto order (N=3 to 20).}
%\label{fig_ph}
%\end{figure}
%The figure shows that convergence can be reached for any $h$ with an increasing Lobatto order.

\begin{figure}[!hbt]
\centerline{\epsfig{file=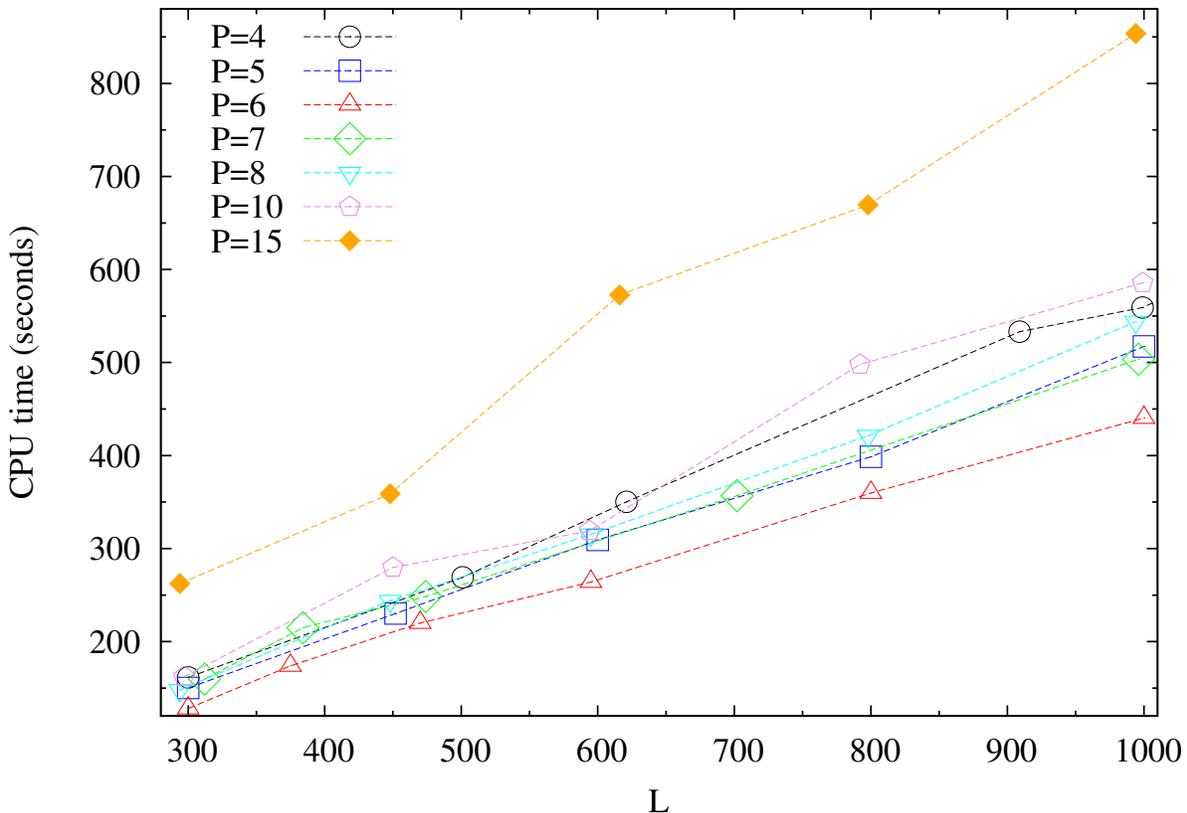,width=.99\columnwidth,angle=0}}
\caption[]{CPU time in seconds for the energy-dependent part of the algorithm as 
a function of number of integration points $L$ for a selection of Lobatto orders $P$ and $M$ elements. 
The data concern the $J=8$ partial wave computations.}
\label{fig_cpu}
\end{figure}

An analysis of the CPU time needed for given accuracies is presented in Tab.~\ref{tab_cpu}.
%\begin{table}[h!bt]
%\caption{CPU (in seconds) for various $N$ and $h$ combinations corresponding to a given value of the error on $\delta$ extracted from fig.~\ref{fig_error}.}
%\label{tab_cpu}
%\begin{ruledtabular}
%\begin{tabular}{c|rr|rr|rr}
% & \multicolumn{2}{c|}{$\log_{10}(\Delta \delta)=-1$} & \multicolumn{2}{c|}{$\log_{10}(\Delta \delta)=-2$}   &  \multicolumn{2}{c}{$\log_{10}(\Delta \delta)=-3$} \\ 
%$N$ & $h$ & CPU & $h$ & CPU & $h$ & CPU  \\ \hline
%3  & 0.18 &1278 & 0.10 & 1435 &      &      \\
%4  & 0.56 & 545 & 0.38 &  795 & 0.26 & 1203 \\
%5  & 1.03 & 369 & 0.77 &  491 & 0.58 &  639 \\
%6  & 1.55 & 279 & 1.23 &  373 & 0.98 &  429 \\
%7  & 2.24 & 281 & 1.81 &  339 & 1.47 &  428 
%\end{tabular}
%\end{ruledtabular}
%\end{table}
%\begin{table}[h!bt]
%\caption{CPU (in seconds) for various $N$ and $h$ combinations corresponding to a given value of the error on $\delta$ extracted from fig.~\ref{fig_error}.}
%\label{tab_cpu}
%\begin{ruledtabular}
%\begin{tabular}{c|rr|rr}
% & \multicolumn{2}{c|}{$\log_{10}(\Delta \delta)=-1$} & \multicolumn{2}{c}{$\log_{10}(\Delta \delta)=-2$}    \\ 
%$N$ & $h$ & CPU & $h$ & CPU    \\ \hline
%3  & 0.18 &1278 & 0.10 & 2267 \\
%4  & 0.56 & 545 & 0.38 &  795 \\
%5  & 1.03 & 369 & 0.77 &  491 \\
%6  & 1.55 & 279 & 1.23 &  373 \\
%7  & 2.24 & 281 & 1.81 &  339 
%\end{tabular}
%\end{ruledtabular}
%\end{table}
\begin{table}
\caption{CPU (in seconds) for various $P$ and $h$ combinations corresponding to two given values of the error on $\delta$ extracted from fig.~\ref{fig_error}. The columns $E_{col}$ ind. and $E_{col}$ dep. correspond to the CPU time of steps to be done once per collisional energy ($E_{col}$ ind.) and at each collision energy ($E_{col}$ dep.).}
\label{tab_cpu}
\begin{ruledtabular}
\begin{tabular}{c|ccc|ccc}
 & \multicolumn{3}{c|}{$\log_{10}(\Delta \delta)=-2$} & \multicolumn{3}{c}{$\log_{10}(\Delta \delta)=-6$}    \\
type    & $h$  & $E_{col}$ ind. &  $E_{col}$ dep. & $h$ & $E_{col}$ ind. &  $E_{col}$ dep.  \\ \hline
$P=3$   &  0.10 &  368   & 2267 &  & & \\   %0.01 &      &      \\
$P=4$   &  0.38 &  140   & ~795 &  & & \\   %0.08 &      &      \\
$P=5$   &  0.77 &   95   & ~492 &  0.24 &  282 &1556  \\
$P=6$   &  1.23 &   73   & ~373 &  0.49 &  180 & ~814  \\
$P=7$   &  1.81 &   59   & ~339 &  0.80 &  141 & ~778  \\
Johnson &  0.10 &  354   & ~~80 &  0.01 & 3422 & ~786  \\
% raw data below but in article this is 551*2/3.  etc
%$P=3$   &  0.10 &  551   & 2267 &  & & \\   %0.01 &      &      \\
%$P=4$   &  0.38 &  187   & ~795 &  & & \\   %0.08 &      &      \\
%$P=5$   &  0.77 &  119   & ~492 &  0.24 &  353 &1556  \\
%$P=6$   &  1.23 &   88   & ~373 &  0.49 &  216 & ~814  \\
%%$P=7$   &  1.81 &   69   & ~339 &  0.80 &  165 & ~778  \\
%Jonhson &  0.10 &  354   & ~~80 &  0.01 & 3422 & ~786  \\
%type    & VRI & VRK \\
%Jonhson & 3422 & 786 \\
%$P=7$   &  165 & 778 \\
%$P=6$   &  216 & 814 \\
%$P=5$   &  353 &1556 \\
\end{tabular}
\end{ruledtabular}
\end{table}
The table presents as a function of the Lobatto order $P$ the CPU in
second needed to reach a given accuracy of $\log_{10}(\Delta \delta)$. The
corresponding $h$ are also listed in the table.  For these calculations
we increase $J$ to 10 for which the number of channels increases to
565 channels with 129 of them energetically open. The CPU given in the
table corresponds to the resolution of the equations for one value of
the collisional energy after an initialization step which is energy
independent and thus to be performed only once if multiple collisions energies are considered.
%which takes for example 296 seconds for $N=3$, $h=0.18$ and 56 s for $N=7$, $h=2.24$.
The table shows that for a given accuracy, one gains in increasing the
Lobatto order at least to the tested orders. The comparison of the
CPU time needed by the Johnson~\cite{1973-BRJ-JCoP-445}
log-derivative propagator is clearly in favor of this last approach
when low accuracy is required.
%with CPU times of 46 and 80 seconds for the two levels of accuracy tested.  
However the situation changes when
high accuracy is needed. In the present test, a $P=7$ computation is
always more efficient than the Johnson version even for a
single collision energy. When using the Gauss-Lobatto discretization,
the CPU time requirement of the collisional energy dependent step is
closely related to the number of integration points $L$ as underlined
by Fig.~\ref{fig_cpu}.
A roughly linear dependence of the CPU time as a function of the number
of integration points is found for the Lobatto orders we tested. At a
given number of integration points, the general trend is an increase of
the CPU time with the Lobatto order.  Some exceptions are found like the
$P=6$ case presented in Fig.~\ref{fig_cpu} which turns out to be cheapest
calculation with respect to  CPU time for all numbers of integration
points from 300 to 1000.  We infer that this is due to particularities
in the sparseness structure of the matrices handled by PARDISO. Similar
but less marked exceptions have been found for the $J=10$ and $J=14$ partial wave
computations.
\subsection{Bound states}
With the appropriate boundary conditions built in Eq.~\eqref{eq_bs}, bound
states of the triatomic Rb$_2$He can be determined using the same
discretized Hamiltonian. We solve the sparse eigenvalue problem Eq.~\eqref{eq_bs}
using the density-matrix-based algorithm FEAST, a package included in
the MKL library based on a contour representation of the resolvent in the complex
plane~\cite{2009-EP-PRB-115112}. Internally, FEAST solves a series
of sparse linear systems using a user-defined subroutine, PARDISO in
our case.

\begin{figure}[!hbt]
\centerline{\epsfig{file=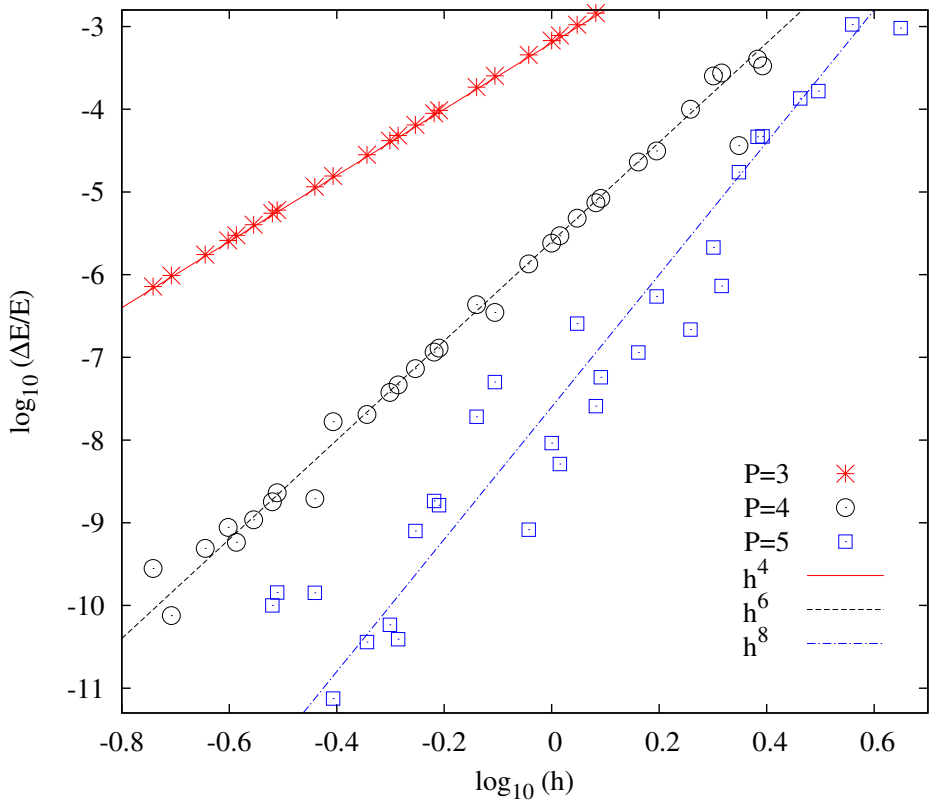,width=.79\columnwidth,angle=0}}
\caption[]{Error as a function of element size for various Lobatto orders ($P=3, 4,$ and 5) together with relevant power lines in log-log scale.}
\label{fig_bound}
\end{figure}

We focus on the $J=2, \Pi=-$ partial wave for which a single
bound state is found. This state is weakly bound with respect to the
Rb$_2$ + He asymptote and an enlarged $R$ box with $R\in [4,236]$~$a_0$
is used. The converged energy is -12.63~milliK below the Rb$_2$ + He
asymptote. Fig.~\ref{fig_bound} presents the evolution of the relative
error on the computed energy when increasing by two the element size $h$
for various  Lobatto order $P$ in log-log scale.
For the lowest $P$ orders presented, the $h^{2P-2}$ behavior is retrieved.
For $P=5$ numerical noise increases when reducing $h$. This is partly
due to the FEAST algorithm which implies an iterative procedure with
two kinds of internal convergence criteria. One criterium tests the
evolution of the energies from one iteration to the next one and the
second one is a maximum number of iterations. The data presented have
been obtained with a $10^{-14}$ value and 50 iterations maximum for
these two FEAST parameters.

\section{Conclusions}

In summary, we have explored the numerical performance of the
spectral-element method in multichannel quantum dynamics. Combination
of the spectral-element discretization with purely diabatic or
diabatic-by-sector bases leads to a highly sparse representation of
the Hamiltonian.
This results in significant memory saving for the bound
state problem as compared for instance to the scaled DVR
approach~\cite{TiesingaPRA98,KokoulineJCP99}. Regarding the scattering
problem, accuracy is significantly less limited by round-off errors in
the spectral element approach than in popular propagation methods and
the corresponding computation time is advantageous when the required
accuracy is high.

In perspective, it may be interesting to test iterative rather than
direct algorithms to solve the scattering linear system for the
discretized Schr{\"o}dinger equation, in particular when boundary
conditions of Eq.~\eqref{eq_smat} are imposed to obtain a single column
of the scattering matrix. In this case, if iterative solvers turned out
to perform better than (number of nonzero elements)$^{3/2}$ one might
be able to overcome the (number of channel)$^3$ unfavorable
computational cost scaling presented by time-independent calculations
as compared to time-dependent calculations.

\begin{acknowledgments}
This work is supported by the Agence Nationale de la Recherche (Contract COLORI No. ANR-12-BS04-0020-01).
\end{acknowledgments}

%\bibliography{Journal,Add}


\end{document}